# Scalable functionalization of optical fibers using atomically thin semiconductors

Gia Quyet Ngo[1], Antony George[2*], Robin Tristan Klaus Schock[1], Alessandro Tuniz[3], Emad Najafidehaghani[2], Ziyang Gan[2], Nils C. Geib[1], Tobias Bucher[1], Heiko Knopf[1,4,7], Sina Saravi[1], Christof Neumann[2], Tilman Lühder[5], Erik P. Schartner[6], Stephen C. Warren-Smith[6], Heike Ebendorff-Heidepriem[6], Thomas Pertsch[1,4,7], Markus A. Schmidt[5], Andrey Turchanin[2*], Falk Eilenberger[1,4,7*]

1. Institute of Applied Physics, Abbe Center of Photonics, Friedrich Schiller University, Albert-Einstein-Str. 15, 07745 Jena, Germany

2. Institute of Physical Chemistry, Abbe Center of Photonics, Friedrich Schiller University Jena, Helmholtzweg 4, 07743 Jena, Germany

3. University of Sydney Nano Institute (Sydney Nano), School of Physics, Physics Road, Camperdown NSW 2006, Australia

4. Fraunhofer-Institute for Applied Optics and Precision Engineering IOF, Albert-Einstein-Str. 7, 07745 Jena, Germany

5. Leibniz Institute of Photonic Technology (IPHT), Albert-Einstein-Str. 9, 07745 Jena, Germany

6. ARC Centre of Excellence for Nanoscale BioPhotonics (CNBP), Institute for Photonics and Advanced Sensing, School of Physical Sciences, University of Adelaide, Adelaide SA 5005, Australia

7. Max Planck School of Photonics, Germany

* Corresponding Authors: Falk Eilenberger (falk.eilenberger@uni-jena.de), Antony George (antony.george@uni-jena.de), Andrey Turchanin (andrey.turchanin@uni-jena.de).

**Atomically thin transition metal dichalcogenides are highly promising for integrated optoelectronic and photonic systems due to their exciton-driven linear and nonlinear interaction with light. Integrating them into optical fibers yields novel opportunities in optical communication, remote sensing, and all-fiber optoelectronics. However, scalable and reproducible deposition of high quality monolayers on optical fibers is a challenge. Here, we report the chemical vapor deposition of monolayer $MoS_2$ and $WS_2$ crystals on the core of microstructured exposed core optical fibers and their interaction with the fibers´ guided modes. We demonstrate two distinct application possibilities of 2D-functionalized waveguides to exemplify their potential. First, we simultaneously excite and collect excitonic 2D material photoluminescence with the fiber modes, opening a novel route to remote sensing. Then we show that third harmonic generation is modified by the highly localized nonlinear polarization of the monolayers, yielding a new avenue to tailor nonlinear optical processes in fibers. We anticipate that our results may lead to significant advances in optical fiber based technologies.**

The light matter interaction length in monolayer transition metal dichalcogenides (TMD) [1, 2] on planar substrates is restricted to sub-nanometres due to their miniscule thickness. This limitation reduces the total optical response of the TMDs and restricts possible applications severely. Therefore, strategies which enhance the light matter interaction are highly desired. Coupling the TMDs with different types of optical resonators is a widely used method to enhance the light matter interaction [3-8]. It is, however, naturally limited to narrowband resonances, while broadband, ultrafast operation cannot easily be implemented. On the other hand, by integrating TMDs on waveguides or optical fibers the interaction length can be enhanced greatly in a broadband, non-resonant manner [9]. The resulting 2D-functionalized waveguides (2DFWG) can utilize the optical properties of TMDs via interaction with parts of the evanescent fields of the guided modes. 2DFWGs can show remarkable features, leveraging from, e.g., the nonlinear or excitonic properties of the TMDs. Previous attempts to fabricate 2DFWG rely on mechanical transfer of exfoliated TMDs onto the waveguides or optical fibers [10-14]. However, this approach is prone to induce uncontrollable stress fields and lacks reproducibility and scalability. Thus, such methods are hardly suitable for future large-scale integration. A process to grow high quality monolayer TMDs directly on optical fibers or waveguides is therefore required to establish 2DFWGs as a new photonic platform. We tackle this challenge by directly growing monolayer TMDs on the guiding core of all-silica microstructured exposed core optical fibers (ECFs) [15], a cross sectional SEM-image of which can be found in Fig. Supp. 2 (a) and (b). Details on ECF-fabrication are given in the Methods section. The growth of 2D-materials turns the ECFs into 2DFWGs, in a scalable process. Specifically, we show the growth of monolayer $MoS_2$ and $WS_2$ crystals on the guiding core of all-silica ECFs and investigate their interaction with the evanescent fields of highly confinement guided modes. The growth is facilitated by a modified chemical vapour deposition (CVD) process [16]. ECFs are compatible with this process, because they consist entirely of silica, which is a well-studied substrate material for high-temperature CVD processes and in particular for the growth of TMDs. The process yields interspersed monolayer TMD crystals of high quality with a typical length of 20 $\mu m$ on ECFs with a length of a few centimetres. Our scalable technique paves the way for 2DFWGs as a new tool for integrated optical architectures, active fiber networks, nonlinear light sources, distributed sensing, and photonic chips.

We highlight the possible functionalities of our 2DFWGs in two case studies. The first demonstrates in-fiber excitation and collection of exciton-driven photoluminescence (PL) which may pave ways towards future experiments in excitonics, remote fiber-based sensing schemes and surface-sensitive bioanalytics. The second is focussed on the way the highly nonlinear TMD coating modifies the nonlinear wave dynamics in ECFs, by investigating enhanced third harmonic generation (THG). In general, this shows that 2DFWGs can be used to enhance and tailor the nonlinear response of integrated wave systems, without any modification to the guided modes themselves, leading to new applications in nonlinear light conversion and optical signals processing. The overall concept of both of our experiments is displayed in Fig. 1 (a). The ECFs have been coated with $MoS_2$ and $WS_2$ crystals on the entire grooved surface, which also forms the upper surface of the ECF's core. A laser is coupled into the fundamental mode (FM) of the ECF, which interacts with the TMDs via the evanescent part of the mode. The resulting polarization; e.g., PL or third harmonic (TH) light, is coupled back into the fiber modes or into free space, where it can be collected for further analysis. An optical microscopy image of the coated exposed side of the ECF is given in Fig. 1 (b), showing high quality $MoS_2$ crystals. The focal plane of the image is chosen to coincide with bottom of the groove running along the entire 60 mm length of the coated ECF, which is also the top of the exposed core.

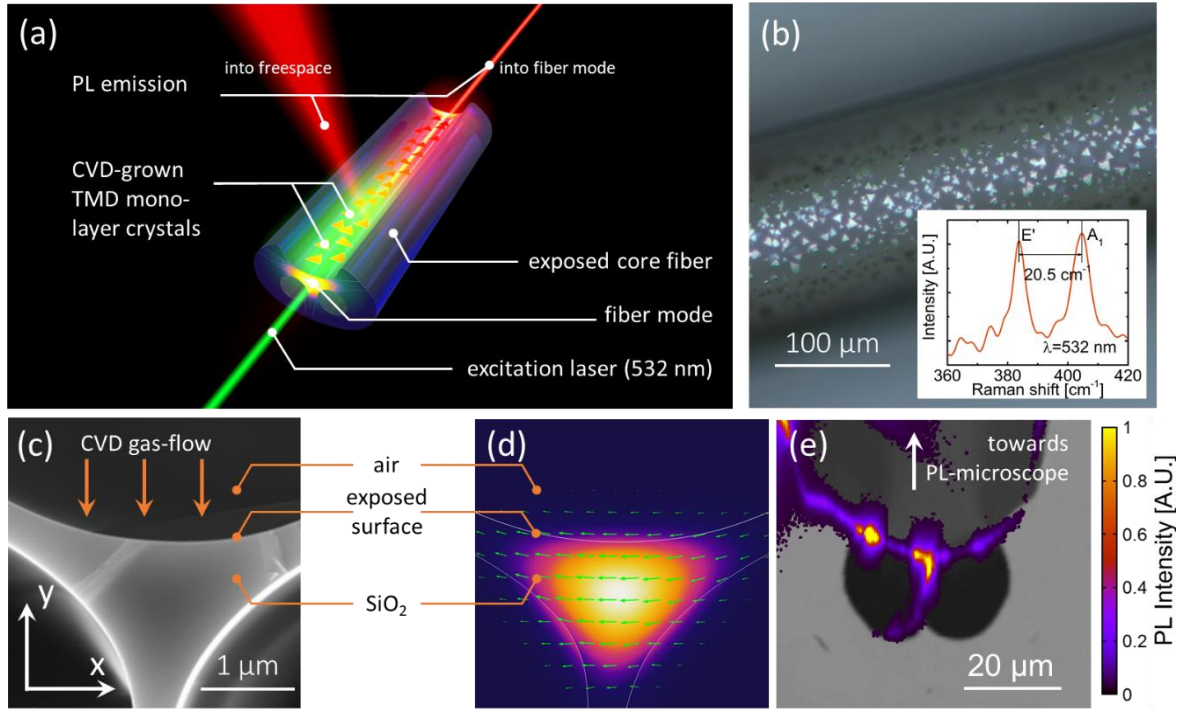

*Fig. 1: (a) Artist's impression of a PL experiment with an ECF based 2DFWG, where PL is excited with the fiber mode and is emitted into freespace and into the fiber mode which can be detected in either way. (b) Optical microscope image of the MoS$_2$ coated ECF. The MoS$_2$ crystals on the exposed core section of the fiber are clearly visible as bright triangles. The inset show typical Raman spectrum of MoS$_2$ monolayer crystals grown on an ECF. (c) Cross sectional SEM image of the core area of the ECF. Bright regions are SiO$_2$, whereas dark regions are free of material. The gas flow on the CVD reactor is marked by orange arrows (d) Simulated norm of the electric field distribution of the fundamental guided mode at 1570 nm. (e) Cross sectional PL mapping of MoS$_2$ coated ECF superimposed with a SEM image for clarification.*

Fig. 1 (c) displays a cross sectional scanning electron microscopy (SEM) image of the ECF's core area (an SEM image of the entire ECF cross section is provided in Fig. Supp. 2 (a)). The core is suspended by three struts of silica to a homogenous silica cladding structure. The upper boundary of the core forms the bottom of a groove, which is running down the entire length of the ECF. When placed in the CVD reactor [16] the upper boundary of the core is thus completely exposed to the CVD reactants. Thus, as monolayer TMD crystals are grown on the entire outer surface of the ECF, they are also grown on the exposed upper surface of the core. Their lateral size, distribution and thickness can be tuned in the growth process. After careful optimization, monolayers were grown almost exclusively, as can be seen from the inset of Fig. 1 (b), which displays a typical Raman spectrum of the MoS$_2$ crystals [17] on the ECF with a characteristic spacing of 20.5 cm$^{-1}$ between the Raman modes. This spacing is consistent with that expected for CVD-grown monolayer TMDs [16, 17]. Examples of alternative growth modes together with their Raman spectra are displayed in Fig. Supp. 2 (c-e). For more information on the growth process, please see the Methods section.

The ECF's core has a diameter of ~2 μm and supports two nondegenerate FMs, which are mostly polarized along the x and y direction. However, the x-polarized FM, with the polarization aligned parallel to the coated surface, has a better field overlap with the TMD layer (see Tab. Supp. 1) and its polarization is aligned with the large $\chi^{(3)}_{xxxx}$-components of the TMDs nonlinear tensor [18]. Hence, all experiments and simulations are carried out with the x-polarized FM. Its field distribution was calculated numerically (more details in Supplement 6) and is shown for a wavelength $\lambda_0 = 1570$ nm in Fig. 1 (d). Because of the small size of the core, the FM is well-confined. A fraction of 1.6 % of the electromagnetic energy is flowing in the air region above the ECF's core, which can thus interact effectively with the TMD crystals.

## Guided-wave Photoluminescence

Next, we verify the location and PL activity of the TMD crystals, grown on the curved ECF core by performing a cross sectional PL emission mapping. The PL map displayed in Fig. 1 (e) is superimposed on a cross sectional SEM image of the ECF for easier understanding. Illumination and collection of PL light were performed sideways through the groove of the uncut ECF along the direction indicated by the arrow in Fig. 1 (e). The PL follows the outline of the ECF grove, which indicates that the TMD crystals have grown in direct contact with the entire surface of the ECF. Note that the part of PL light extending downwards from the center of the image is caused by the diffraction at the ECF core and its interaction with the confocal setup and does not indicate the presence of TMDs within the core. A PL map along the propagation direction and a PL spectrum can be found in Fig. Supp. 4.

We now focus on guided wave excitation of PL in 2DFWGs, which is mediated by excitation and decay of excitons in the TMD coating. Excitons in TMDs [19] are particularly appealing because they exhibit spin valley coupling [20-23] and are important for the emission of single photons [24-28]. We couple a green laser ($\lambda = 532$ nm) without polarization control into the ECF, which excites excitons via the evanescent field of the FM (for experimental setup see Fig. Supp. ). PL from the TMD monolayers is either emitted into free space or coupled back into the ECF's mode. Emission into guided modes has been observed by imaging the end facet of the fiber and by measuring the spectrum. The results are displayed in Fig. 2 (a) and (b). The image of the PL at the end facet of the ECF, is long-pass filtered to remove residual laser radiation and then displayed in Fig. 2 (a). For better orientation, a microscope image of the ECF itself is superimposed. The PL light is clearly emitted from guided modes at the core of the ECF. This light is then analysed spectroscopically, revealing exciton peaks at $678$ nm ($MoS_2$) and $622$ nm ($WS_2$) and a spectral full-width at half-maximum of 48 nm and 43 nm, respectively. These values do not significantly differ from the ones reported on planar substrates grown by the same technique [16]. We thus conclude the material properties are comparable to those grown on planar substrates and no additional strain has been induced, e.g. due to different thermo-mechanical properties of the substrate.

Observing PL in the guided mode of the fiber means that the evanescent field of guided modes can be used to both excite and collect PL [29-31] in an integrated optical environment. This makes 2DFWGs, such as TMD coated ECFs, highly interesting for integrated excitonics and remote sensing applications.

Lateral emission into free space was observed with a camera mounted sideways, imaging the bottom of the ECFs groove. We attain compound images of the distribution of PL over a substantial section of the ECF and thus an image of PL active TMD crystals. One image such of an $MoS_2$ coated fiber is displayed in Fig. Supp. 5 (b). From this, we can extract the distribution and cumulative length of monolayer crystals on the ECF. For this specific sample we observe 39 distinct $MoS_2$ crystals with an average length of $29$ μm per crystal and a filling factor of $5.4\%$, although the coverage and crystal size have been significantly increased in later batches after further optimization of our growth procedure.

A transmission spectrum through the ECF is displayed in Fig. 2 (c). It was obtained by coupling white light into the $MoS_2$ coated ECF. This spectrum is shown with the imaginary part of the refractive index of $MoS_2$, the two absorption peaks at $619$ nm and $671$ nm can thus be clearly connected to the characteristic exciton resonances the TMD.

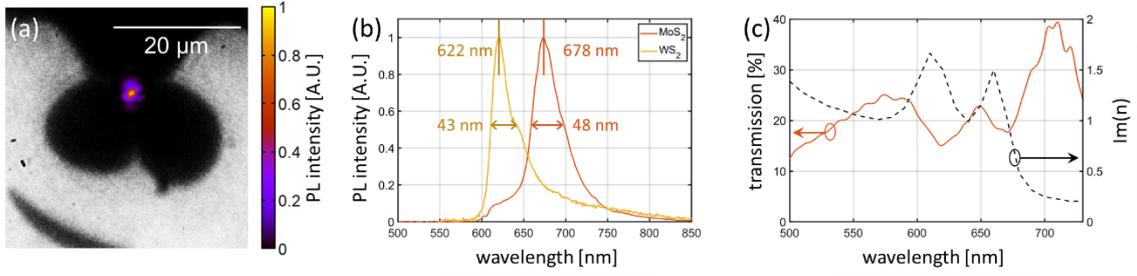

*Fig. 2: Modal PL and transmission measurement, where the laser/white light source is coupled into the TMD coated ECF's guided mode, excites excitons in the TMD coating, which emit PL back into the fundamental mode. The PL signal is then recorded after it has propagated through the fiber. (a) False color image of PL excited by a 532 nm laser in an MoS$_2$ coated ECF. The background image is the back surface of the ECF, recorded through the same camera. (b) Normalized PL spectra of MoS$_2$ and WS$_2$ coated ECF excited by a 532 nm laser. (c) Transmission spectrum through an MoS$_2$ coated ECF. A LED white light source is used for transmission measurement. Values $\lambda > 700\ nm$ are highly influenced by measurement inaccuracy, due to a lack of power of the light source for this spectral region.*

## Enhancement of Third Harmonic Generation

2D TMDs are also highly interesting because of their strong nonlinear optical response per unit thickness [32]. For third order processes this is quantified by the nonlinear refractive index $n_2$ with a reported value of $n_2^{\text{MoS}_2} \sim 2.7 \cdot 10^{-16}\ \text{m}^2/\text{W}$ for TMDs transferred on waveguides [14]. It is almost four orders of magnitude larger than that of silica, although lower values have been reported on planar substrates [18]. Thus, a TMD coating may have a substantial contribution to nonlinear effects in ECFs, although less than $10^{-4}$ of the power flow of the FM is localized in the TMD at any wavelength (see Fig. Supp. 6 (e)). The influence on the nonlinearity can be quantified by calculating the respective contributions of the MoS$_2$ coating and the SiO$_2$ core to the overall self-phase modulation coefficient $\gamma = \gamma_{\text{MoS}_2} + \gamma_{\text{SiO}_2}$ [33]. Indeed, we find $\gamma_{\text{MoS}_2} > \gamma_{\text{SiO}_2}$ for wavelength in excess of 1470 nm (see Supplement 7 and Fig. Supp. 6(g)), i.e. the nonlinear contribution of the TMD coating dominates for long wavelengths. Note, that even larger $\gamma_{\text{MoS}_2}$ may be obtained in the future by optimizing the field overlap of the FM with the TMD coating, opening e.g. the path for TMD enhanced supercontinuum generation experiments.

While many third order nonlinear processes are observed in fibers, THG is particularly fascinating, because it relies on the simultaneous interplay of nonlinearity, mode matching, and phase matching (PM). We found that there is no appreciable modification of the PM by the TMD coating, because all linear mode properties, except loss, are unaffected by the TMD coating (see Fig. Supp. 6). The ECFs of the design used here, exhibit PM only for higher-order TH modes (HOMs) at a TH wavelength of roughly 550 nm, corresponding to a fundamental wavelength of 1650 nm [34-36].

To show that THG is indeed enhanced we excited TH with a pulsed laser operating at $\lambda_0 = 1570$ nm and a pulse duration of 32 fs (see Fig. Supp. 7 for pulse characterization [37]. Fig. 3 (a) displays the TH spectrum for three different input energies for an uncoated and an MoS$_2$ coated ECF. Measurements were qualitatively reproducible for individual ECFs but also across different samples. For ease of representation we present data here from a representative sample. We consistently observe more THG in the MoS$_2$ coated ECF, which signifies that the TMD coating does enhance the THG process. This is particularly noteworthy, as the MoS$_2$ coated ECFs experience roughly 60% linear loss over the length of the ECF and the comparison was made for equal input energy. As dictated by PM we observe TH not at exactly a third of $\lambda_0$ but in a spectral band ranging from 540 nm to 560 nm, marked in Fig. 3 (a). The fundamental wave (FW) spectrum must thus first nonlinearly broaden (more details in Fig. Supp. 8) into a THG-relevant sub-band between 1620 nm to 1680 nm before TH is generated. This explains the somewhat stronger-than-cubic scaling in the inset Fig. 3 (a). The similarity of TH spectrum for both

ECF types reassures us that the phase matching (PM) between FW and TH is indeed unaffected by the MoS$_2$ coating.

Because a PM modification is thus ruled out, the observed TH enhancement must be related to the process of nonlinear light generation itself. This process is driven by the nonlinear polarization field $\boldsymbol{P}^{\text{THG}}\left(\frac{\lambda_0}{3}\right) = E_x^3(x, y; \lambda_0) \cdot \chi_{xxxx}^{(3)}(x, y)\boldsymbol{e}_x$. Here, $E_x(x, y)$ is the x-component of the electric field of the FW mode and $\chi_{xxxx}^{(3)}(x, y)$ is the dominating element of the nonlinear tensor of the THG process for the TMD coating and the silica core, respectively. The shape of $\boldsymbol{P}^{\text{THG}}$ is displayed in Fig. 3 (b) in logarithmic scaling. There are two major contributions: the spatially smooth nonlinear polarization from the SiO$_2$ core and the strong but highly localized contribution of the TMD coating, better visible in the inset.

The so-generated TH radiation is distributed onto the TH modes, the magnitude of which is described by an overlap coefficient $\gamma_k^{(\text{THG})}$ for every TH mode (see supplement). Thus, the addition of the TMD coating does not enhance the nonlinear interaction for all HOMs equally but it boosts those that are localized close to the surface and with predominant x-polarization. This mode-selective nonlinear enhancement is quantified in Fig. 3 (c), displaying $\gamma_k^{(\text{THG})}$ values for all TH HOMs close to the PM point. The PM region contains 11 HOMs and is marked by the shading (see supplement for determination of PM bandwidth $\Delta n$). The contribution of SiO$_2$ and MoS$_2$ are marked in different colours. While $\gamma_k^{(\text{THG})}$ grows for all HOMs, it does so very differently from mode, i.e. the enhancement is mode-selective.

To confirm this model and to compare with the experiment, we only discuss the three HOMs with the largest $\gamma_k^{(\text{THG})}$ marked with $M_1$ to $M_3$ in Fig. 3 (c). The most dominant HOM is the $M_1$ mode at $n_{M_1}^{\text{eff}} = 1.379$. $\gamma_{M_1}^{(\text{THG})}$ is approximately doubled due to the MoS$_2$ coating. The second strongest contribution for the bare ECF comes from the $M_2$ mode at $n_{M_2}^{\text{eff}} = 1.366$, $\gamma_{M_2}^{(\text{THG})}$ increases by ~2.5 upon application of the TMD coating. Upon coating; however, it is superseded by the $M_3$ mode at $n_{M_2}^{\text{eff}} = 1.383$, which almost quadruples its $\gamma_{M_3}^{(\text{THG})}$. The modes profiles of $M_1$, $M_2$, and $M_3$ can be found in Fig. Supp. 9.

The mode-selective enhancement of the overlap coefficients is reflected in spatial distribution of the TH light as seen in images recorded by a camera, focussed to output plane of the ECFs, displayed in Fig. 3 (d) and (e), for a bare and an MoS$_2$ coated ECF of identical length, respectively. The single peak marked with the white oval at the top of bare the ECF is replaced by a broader and weaker triple peaked distribution for the MoS$_2$ coated ECF. Moreover, the field is less localized and extends further into the bottom strut for the MoS$_2$ coated ECF in the region marked with the green circle. Qualitatively, both distributions can be reproduced by a simple superposition of the $M_1$, $M_2$ and $M_3$ modes, displayed in Fig. 3 (f) and (g), where the superposition coefficients are chosen according to the relative values of the overlap coefficients $\gamma_{M_{1-3}}^{(\text{THG})}$ with and without the MoS$_2$ coating. Both the modification at the top and bottom of the image are well-reproduced for such a simple model, particularly in light of the large uncertainties of the linear and nonlinear coefficients of MoS$_2$ and the impact of the random distribution of the crystals on the ECF.

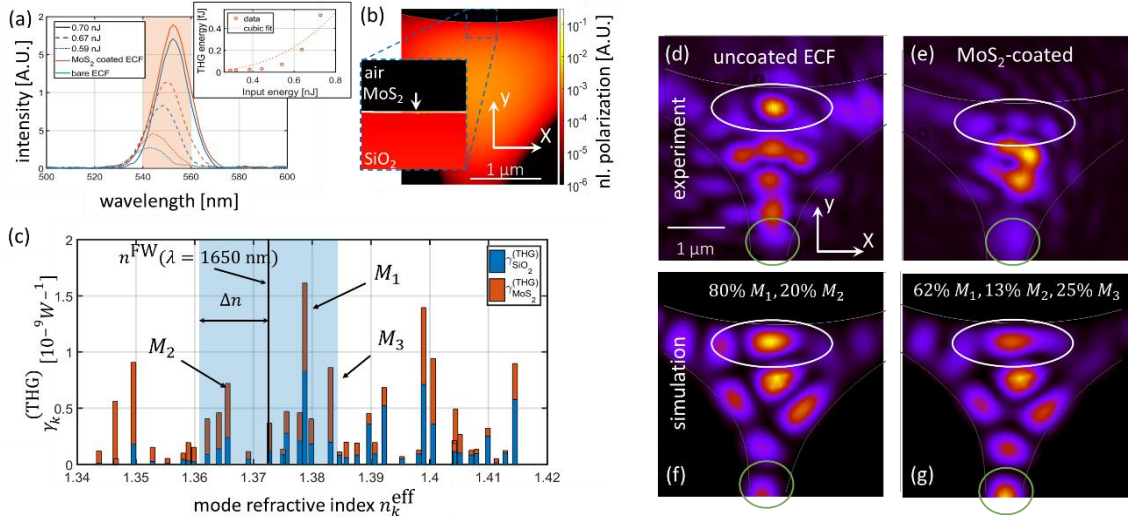

*Fig. 3: (a) Third harmonic spectra. The THG band from 540 to 560 nm is marked in orange. (Inset) Power dependence of the THG for the MoS$_2$ coated ECF, with a cubic fit. (b) Log-Plot of the nonlinear polarization field, driving THG. The field is generated by the cubed FM and the nonlinear action of SiO$_2$ and MoS$_2$. The inset is a zoom of the exposed surface, showing the strong and highly localized nonlinear polarization from the MoS$_2$. (c) Modal THG overlap coefficients $\gamma_k^{(THG)}$ at a TH wavelength of 550 nm, calculated for all HOMs k, which are close to the phase matching point at $n_{eff}^{FW} = 1.373$. The PM region with $\Delta n < 0.012$ is marked in blue. The stacked colored bars mark the contributions of the SiO$_2$ core (blue) and the MoS$_2$ coating to $\gamma_k^{(THG)}$ (orange), respectively. (d) Experimentally acquired image of the light distribution of the THG field at the end of a bare ECF. (e) Experimentally acquired picture of the light distribution of the THG field at the end of an MoS$_2$ coated ECF. (f) Simulated picture at the end of the ECF as a superposition the $M_1$ and $M_2$, with an 80/20 power distribution, according roughly to the overlap coefficients for the bare ECF. (g) Simulated picture at the end of the ECF as a superposition the $M_1$, $M_2$ and $M_3$, according to a 62/13/25 power distribution according roughly to the overlap coefficients for the MoS$_2$ coated ECF. (d-g) White ovals and green circles mark regions, in which the coated and uncoated ECF show appreciable differences.*

## Summary

In summary, we have shown that high quality crystalline monolayer TMDs, e.g. MoS$_2$ and WS$_2$, can be grown directly on the core of microstructured exposed core fibers in a scalable CVD process. This process functionalizes the optical fibers, creating a new platform to investigate and utilize the electrooptic properties of 2D TMDs. Excitonic and nonlinear functionalization is demonstrated in two case studies. First, we excite and collect excitonic photoluminescence from monolayers in the optical fiber, which may give access to, e.g. remote sensing schemes, leveraging on the sensitivity of the TMD excitons to environment. Moreover, it may provide a new platform to investigate excitonic effects over long interaction lengths. A connection with in-fiber electrodes could be used to facilitate novel ultrafast detectors for light. We have also demonstrated that 2D materials modify nonlinear optical processes intricately by investigating the enhancement of third harmonic generation, which is found to be highly mode selective. This may will enhance the design freedom for highly nonlinear guided wave systems and may be utilized in nonlinear fiber devices. Altogether the direct growth of 2D materials on waveguides is opening a novel path towards the scalable and reproducible functionalization of waveguides, fibers, and other integrated optical systems.

## Contributions

FE was the principal contributor to the manuscript and the overall coordinator of the experiments. AG, MAS, ATur, TP, and FE developed the concept and contributed to the overall course of the research. GQN conducted experiments and simulations for both case studies and was supported by RTKS. AG developed the CVD process and adapted it for the ECFs. ATun and SS supported the modelling of the nonlinear interaction coefficients. EN, ZG and CN were responsible for material growth and structural characterization. NCG, TB, HK, and TL were responsible for optical materials characterization and characterization of the optical experiments. ES, SWS and HEH developed the ECFs fabrication technique, provided samples and supported in the modelling of mode properties. All authors contributed to the manuscript.

## Acknowledgements

FE was supported by the Federal Ministry of Education and Science of Germany under Grant ID 13XP5053A. QGN, EN, and SS are supported by the European Union, the European Social Funds and the Federal State of Thuringia as FGR 0088 under Grant ID 2018FGR00088 and FGR 0067, respectively. SS is also funded through the IMPULSE junior researcher promotion program of the Friedrich Schiller University Jena. NCG and AG were supported by the German Research Council as part of the CRC SFB 1375 NOA project B3 and B2. This project received funding from the joint European Union's Horizon 2020 and DFG research and innovation programme FLAG-ERA under a Grant TU149/9-1. TL and MS acknowledge financial support from the German research foundation via the grants SCHM2655/9-1 and SCHM2655/11-1. ATun is the recipient of an Australian Research Council Discovery Early Career Award (project number DE200101041) funded by the Australian Government. This work was performed in part at the OptoFab node of the Australian National Fabrication Facility utilizing Commonwealth and SA State Government funding. We thank Stephanie Höppener and Ulrich S Schubert for enabling our Raman Spectroscopy studies at the JCSM.

## Methods

**ECF fabrication:** The fused silica ECF was fabricated using an ultrasonic drilled silica preform that was then cut open on one side to expose one part of the central section of the fiber. The preform was then caned and inserted into a jacket tube, which is then drawn into an ECF. Active pressurization was used during the draw on the central cane piece, to stabilize the structure [15]. The fiber has an outer diameter of 220 µm (see Fig. Supp. 2a) and an effective core diameter of 2 µm.

**CVD growth of TMDs on fibers:** $MoS_2$ and $WS_2$ crystals were grown on the ECF (fixed on a quartz holder) by a modified CVD growth method in which a Knudsen-type effusion cell is used for the delivery of sulfur precursor [16] from sulfur-powder and metal-oxide powder is used to supply the transition metal. Details of the method are given in [16], from which no modifications were made, except for the usage of ECFs as the substrate. The grown TMDs on the ECFs were initially characterized using optical microscopy (Zeiss Axio Imager Z1.m) and Raman spectroscopy (Bruker Senterra spectrometer operated in backscattering mode using 532 nm wavelength obtained with a frequency doubled Nd:YAG Laser, a 100x objective and a thermoelectrically cooled CCD detector.)

**Microscopic PL mapping and spectroscopy:** Photoluminescence mapping was carried out with a commercial confocal PL lifetime microscope (Picoquant Microtime 200), with an excitation laser operating at 530 nm. The maps, where created by moving the sample along the focus of the microscope's objective, which had a magnification of 64x. The resulting spatial resolution is estimated be in the range of 500 nm. Detection of the PL signal was carried out with an avalanche photodiode. Alternatively, the PL microscope was connected to grating spectrometer (Horiba Jobin Yvon Triax) equipped with a cooled CCD detector to measure PL spectra.

**In-fiber and transverse PL mapping and spectroscopy:** To record PL and transmission spectra, the incoming light was focused into the fiber core. For PL spectroscopy the outcoming light was passed through two of 550 nm long-pass filters and then imaged into a spectrometer (Horiba Jobin Yvon Triax), with a cooled Si-CCD-detector. PL was excited with a 532 nm laser (lighthouse Photonics Sprout), whereas transmission spectra have been excited with a white light diode. Alternatively, the light was imaged onto a CMOS camera (Zyla 4.2 sCMOS), to image the PL at the output facet. The camera was alternatively mounted laterally together with a 10x objective imaging the sideways emission of PL from these crystals. Again, a set of 550 nm long-pass filters was used to reject scattered light from the excitation laser. The camera and the objective had been mounted on a motion stage to map larger sections of the fiber side.

**THG measurements:** Nonlinear experiments were carried out with a femtosecond laser emitting pulsed with a duration of 32 fs at a central wavelength of 1570 nm at a repetition rate of 80 MHz (Toptica FemtoFiber pro IRS-II) and focussed into the ECF with an aspheric lens. Light leaving the fiber was collimated with a microscope objective and coupled into an optical spectrum analyser, to measure the FW and the TH spectra.

# Supplement to:
# Scalable functionalization of optical fibers using atomically thin semiconductors

Gia Quyet Ngo[1], Antony George[2*], Robin Tristan Klaus Schock[1], Alessandro Tuniz[3], Emad Najafidehaghani[2], Ziyang Gan[2], Nils C. Geib[1], Tobias Bucher[1], Heiko Knopf[1,4,7], Sina Saravi[1], Christof Neumann[2], Tilman Lühder[5], Erik P. Schartner[6], Stephen C. Warren-Smith[6], Heike Ebendorff-Heidepriem[6], Thomas Pertsch[1,4,7], Markus A. Schmidt[5], Andrey Turchanin[2*], Falk Eilenberger[1,4,7*]

1. Institute of Applied Physics, Abbe Center of Photonics, Friedrich Schiller University, Albert-Einstein-Str. 15, 07745 Jena, Germany

2. Institute of Physical Chemistry, Abbe Center of Photonics, Friedrich Schiller University Jena, Helmholtzweg 4, 07743 Jena, Germany

3. University of Sydney Nano Institute (Sydney Nano), School of Physics, Physics Road, Camperdown NSW 2006, Australia

4. Fraunhofer-Institute for Applied Optics and Precision Engineering IOF, Albert-Einstein-Str. 7, 07745 Jena, Germany

5. Leibniz Institute of Photonic Technology (IPHT), Albert-Einstein-Str. 9, 07745 Jena, Germany

6. ARC Centre of Excellence for Nanoscale BioPhotonics (CNBP), Institute for Photonics and Advanced Sensing, School of Physical Sciences, University of Adelaide, Adelaide SA 5005, Australia

7. Max Planck School of Photonics, Germany

* Corresponding Authors: Falk Eilenberger (falk.eilenberger@uni-jena.de), Antony George (antony.george@uni-jena.de), Andrey Turchanin (andrey.turchanin@uni-jena.de).

## Supplement 1  Experimental Setup

A sketch of the experimental setup is depicted in Fig. Supp. 1. To record PL, and transmission spectra, the incoming light was focused into the fiber core using a 40x microscope objective. An identical objective was used to collimate the light, leaving the other end of the fiber. For PL the outcoming light was passed through two consecutive of 550 nm long-pass filters, to remove residual laser radiation. It was then refocused into the entrance of a multimode fiber with 200 µm diameter connected to a HORIBA spectrometer, with a cooled Si-CCD-detector. The spectrometer was used to measure both PL and transmission spectra. PL was excited with a 532 nm laser, whereas transmission spectra have been using excitation with a fiber coupled white light diode.

The fiber and all objectives have been mounted on 3-axis translation stages, to optimize light coupling and imaging. A CCD camera was placed behind the setup with a flip mirror, with a tube lens. This way the output facet could be observed, which was used to obtain a good coupling into the fundamental mode for PL and transmission and further used to obtain the modal images of the THG field. It was replaced by an InGaAs-camera to optimize the incoupling for the THG experiments.

A highly sensitive Zyla 4.2 sCMOS camera and a 10x objective was utilized to determine the distribution and size of $MoS_2$ crystals on the fiber, by imaging the sideways emission of PL from these crystals. A set of 550 nm long-pass filters was used in this case to reject scattered light from the excitation laser. The camera and the objective had been mounted on a motion stage to map larger sections of the fiber side.

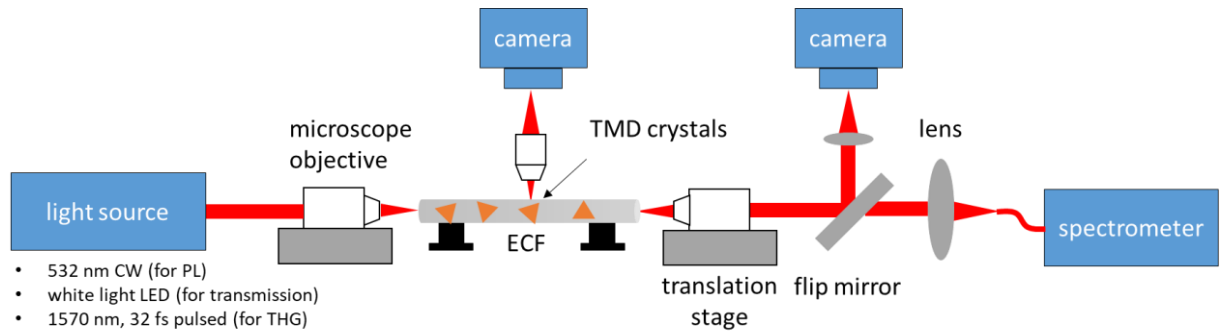

*Fig. Supp. 1: Schematic diagram of experimental setup for PL, transmission and THG measurements. ECF, exposed core fiber samples.*

To demonstrate the THG in the ECF, the pump beam was focused into the ECF using an aspheric lens (Thorlabs C230TMD-C), instead of the microscope objective for better IR focussing. The incident power and polarization were adjusted by a combination of half wave plate and polarizer. The output was imaged using a 40x microscope objective and an IR camera to confirm the coupling into the fiber core. The output was coupled into a graded index multimode fiber (1 mm in diameter) connected to an optical spectrum analyser (ANDO 6315A). The device has a noise level of -80 dBm for the range 350-1750 nm. All signals should be above this detection limit.

## Supplement 2   Fiber Geometry and Raman Spectra of TMD on ECFs

ECFs are microstructured optical fibers of the suspended core type, i.e. they consist exclusively of glass with air insertions that run along the length of the fibers. The air insertions form a silica core, which is suspended by three (or more) thin glass struts. ECFs [1] are a special subclass of suspended core fiber in which the outer wall is removed for one of the air holes so that the entire face of the fiber is exposed to the external environment, i.e. it forms a trench which runs along one side of the fiber. SEM images of the fiber cross section and a zoom into the core area are depicted in Fig. Supp. 2 (a) and (b), respectively.

The ECFs have been placed in a CVD chamber with the trench being exposed to the gas flow of the reactor [2]. As a result, TMD crystals are grown along the entire surface of the fiber, and also in the trench. Some of the crystals grow on the exposed side of the core and can thus interact with the fibers modes. Depending on the specific growth conditions, both single layer, as well as multi layer crystals have been observed after completion of the growth process. Images of such crystals are displayed in Fig. Supp. 2 (c) and Fig. Supp. 2 (e), together with Raman spectra, which underline the respective nature of the two types of crystals in Fig. Supp. 2 (d). Note, that by optimization of the growth parameters, we were able to fabricate fibers almost exclusively coated with single layer crystals.

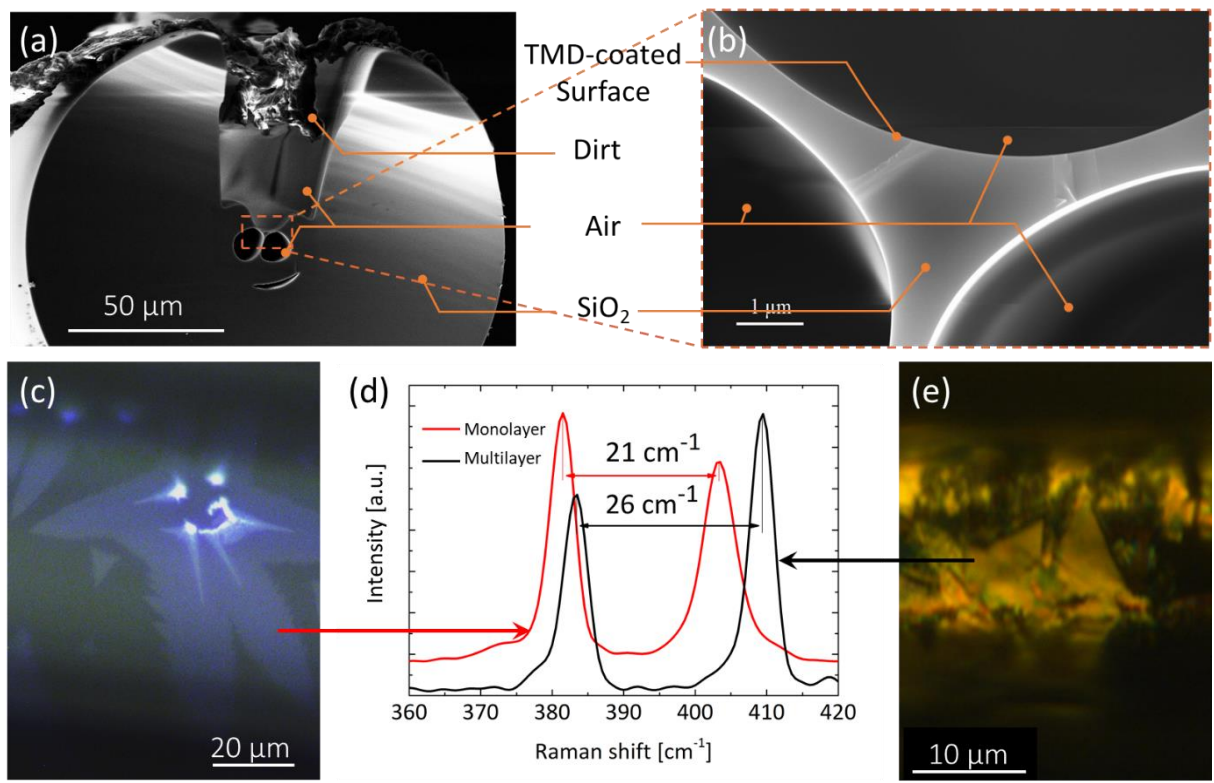

*Fig. Supp. 2: (a) SEM cross section of the entire an ECF. The groove running along the length of the fiber is visible at the top. (b) SEM cross section of the core area of the ECF located at the orange box in (a). (c) Microscope image of a monolayer MoS$_2$ crystal CVD grown on the exposed surface of the core of an ECF. (d) Raman spectra of the MoS$_2$ crystals displayed in subfigures (c) and (e). (e) Microscope image of a multilayer MoS$_2$ crystal CVD grown on the exposed surface of the core of an ECF.*

## Supplement 3    AFM Mapping

We have mapped the morphologic structure of an MoS$_2$-flake grown on the exposed side of the ECF's core with an AFM. Mapping was carried out over a range of roughly 10x10 µm. The resulting topology map is displayed in Fig. Supp. 3 (a). An TMD-crystal with its typical trigonal symmetry is clearly visible in the center of the image. Note that the image contains many contaminations, which we attribute to the fact, that the specific sample under test had been in use in the experiment for months, before AFM was actually measured. We thus assume that it is a consequence of a-posteriori contamination and not related to the growth process.

Fig. Supp. 3 (b) displays a height profile across the boundary of the flake along the line marked in Fig. Supp. 3 (a). The height of the flake was determined to be 0.7 nm, which is a further proof that monolayer TMD-crystals have indeed grown on the ECF.

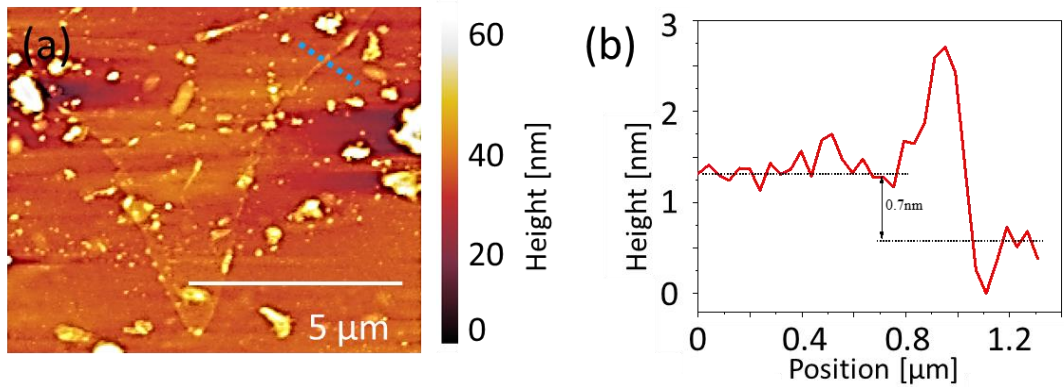

*Fig. Supp. 3: AFM images of monolayer MoS$_2$ on fiber. (a) AFM topography map. (b) AFM height profile as shown by the dashed blue line in (a).*

## Supplement 4   External PL Mapping and Spectra of MoS$_2$ on an ECF

PL was initially tested with a commercial confocal PL microscope using a 530 nm laser for excitation. The microscope has a spatial resolution in order of 500 nm. The fiber was placed flat on the specimen table and thus perpendicular to the optical path. An x-y-map (parallel to the specimen table and along the fiber) was first measured to determine the location of a PL active monolayer crystal. It is displayed in Fig. Supp. 4 (a).

Due to the confocality of the setup and the curved nature of the ECF's core surface, we only map PL from material close to the ECF core itself. Spectra were then measured from various spots along the length of the fiber, with a typical result recorded below. A double gaussian was then fitted for the MoS$_2$ spectrum, as done in Fig. Supp. 4 (b) to judge the relative contribution of the two exciton families observed in MoS$_2$ to the overall PL. Overall, it was found that both width and their relative strength of the peaks all coincide with the values for high quality CVD grown TMDs on planar surfaces.

We then measured a cross section PL map, by scanning perpendicular to the fiber. The results are reported in the main text of the paper. We note that the TMDs have indeed grown conformally to the curved inside surface of the ECF, with no appreciably loss of PL quality confirmed by the measurement of further PL spectra.

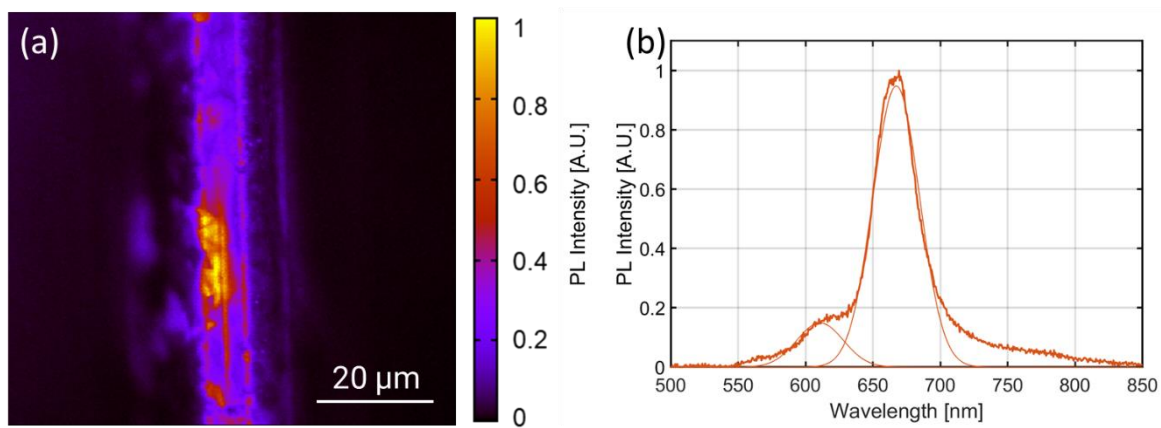

*Fig. Supp. 4: (a) Top view PL mapping of a section of a MoS$_2$ coated ECF. (b) PL Spectrum of the brightest part of the image (bold) together with a double Gaussian fit to the two excitons commonly encountered in MoS$_2$ (thin).*

## Supplement 5   Determination of the Distribution and Size of MoS$_2$ crystals on a Fiber Section

Using the technique described in the experimental setup we could image PL light emanating from the ECF sideways with a resolution of approximately 10 µm using a 10x microscope objective. Translation of the objective and camera with respect to the fiber allowed us to ascertain a set of images for a substantial part of the fiber. Mechanical limitations however prevented us from imaging the entire length of the fiber, particularly the first and the last part of the fiber are difficult to access, due to the mechanical space required by the incoupling and outcoupling objectives.

After recording of the images a compound PL image, such as the one displayed in Fig. Supp. 5 (b) was generated. Note that the image has a different scaling across and along the fiber and is split into 5 subsections for ease of display. A statistical analysis of the acquired dataset was then carried out. A threshold just above the noise level of the camera was set and each connected spot with values above this threshold was considered to be a PL active monolayer crystal. Using the data, we could map both the total light emitted by each crystal together with their size (e.g. length), as displayed in Fig. Supp. 5 (c). The linear connection between the two quantities suggests that each crystal has a comparable brightness per unit length and thus we conclude that their material quality does not change from crystal to crystal.

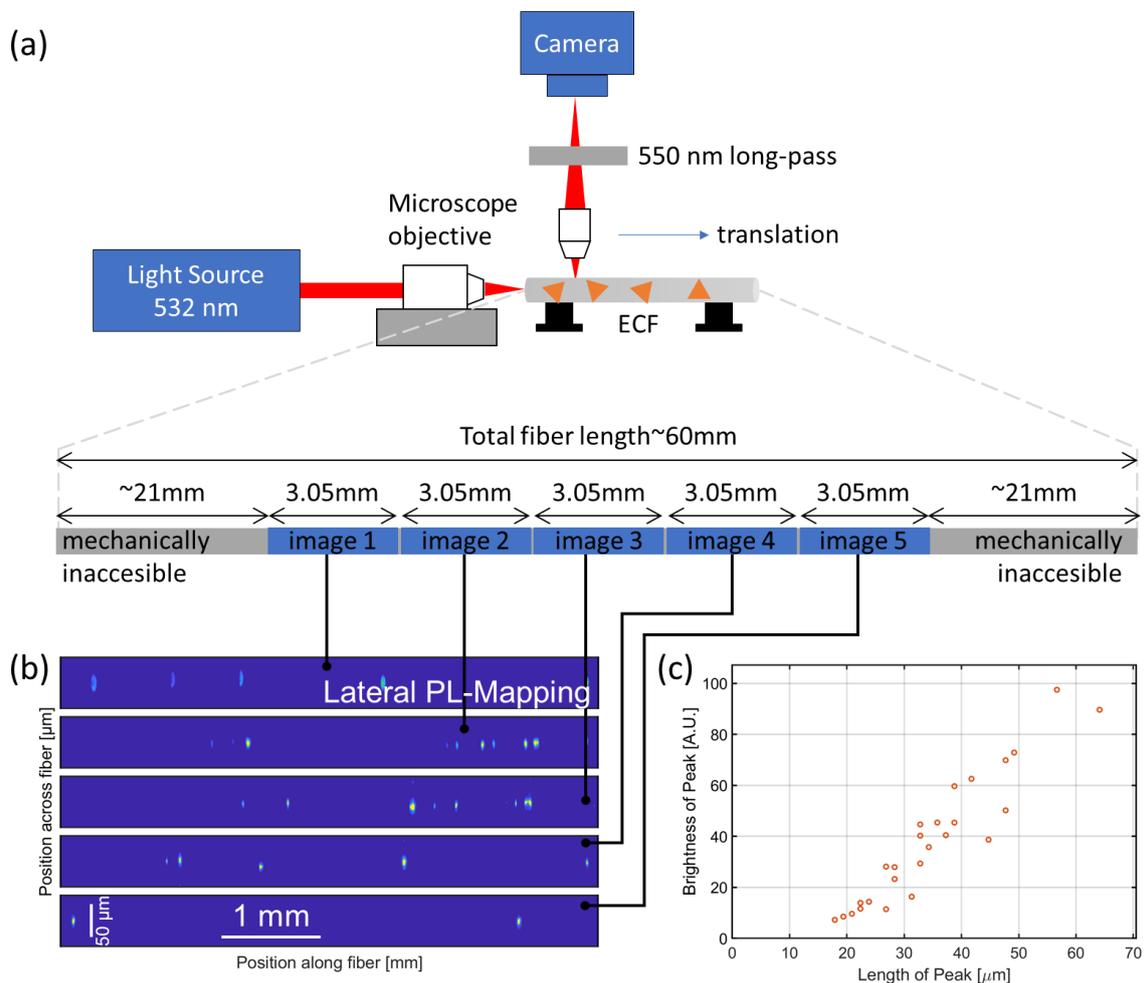

*Fig. Supp. 5: Sideways PL-characterization of TMD-crystal distribution and crystal size. (a) Scheme of the experimental setup together with a sketch of the areas of ECF, which could (blue) and could not (gray) be accessed by the technique. (b) Compound PL side view image of the accessible 18.3 mm section of MoS$_2$ coated optical fiber, when the core is excited with a 532 nm laser, showing the distribution of MoS$_2$ crystals being excited by the fiber mode. The complete image was composed of subimages, which are presented over five lines for the ease of presentation. Analysis of the images reveals 39 distinct peaks with a total length of 0.99 mm for a coverage of 5.4 %. (c) Comparison of the length of the measured PL peaks in (b) and their respective brightness. A more or less linear behaviour shows that all peaks have equal brightness per length, indication homogenous deposition and excitation. The average peak length is 29.1 µm.*

## Supplement 6 Properties of the ECF's Fundamental Mode

Dispersive and nonlinear properties of the coated sections of the ECF were calculated from the geometry measured by the SEM displayed in Fig. Supp. 2 (a) and (b) with a high-index layer of a thickness of 0.65 nm superimposed onto the upper surface of the ECF core. Modal properties have been obtained using finite element simulations (COMSOL v5.4). Scattering boundary conditions were applied to analyse FM and HOMs. As a result of not fully symmetric geometry, the first two modes are non-degenerate and almost mutually orthogonally polarized. Power flow distributions, i.e. the z-component of the Poynting vector, of the fundamental mode is displayed at three different wavelengths in Fig. Supp. 6 (a-c).

The refractive index of $SiO_2$ was calculated directly by COMSOL as a function of frequency using the Sellmeier equation. The refractive index of $MoS_2$ was taken from [3], as displayed in Fig. Supp. 6 (f) while the refractive index of air was set to be 1.0. In this simulation, the ECF core was established as a lossless material, so the damping of the travelling light, as displayed in Fig. Supp. 6 (d)) is caused by the high extinction coefficient of $MoS_2$ monolayer.

Propagation loss was also confirmed experimentally. For the bare ECFs, we achieve a coupling efficiency of 29% at $\lambda = 1570$ nm, with no appreciable propagation loss (less than 1 dB/m) [4]. For the $MoS_2$ coated ECF, we assume the same coupling efficiency. However, the transmitted power is roughly 60% lower at otherwise identical settings. The total fiber length was 60 mm, so that we measure an overall net loss of 0.1 dB/mm. Using the filling factor of 5.4% established in Supplement 5, this means that the $MoS_2$ coated sections of the ECF are responsible for a loss of ~1.8 dB/mm. This value is fairly consistent with the simulated values in Fig. Supp. 6(d), which predict a value of 1.2 dB/mm and an independent cut back measurement, which yielded 1.3 dB/mm.

Fig. Supp. 6 (d) also displays the modal dispersion, which is an important characteristic to describe the spectral broadening of the fundamental wave due to soliton induced supercontinuum generation. The modal dispersion is identical within the boundaries of the precision of the solution with the TMD coating, which is a direct indication that the miniscule coating of the high index TMD does not affect the linear properties of the fundamental mode.

The fraction of power in the TMD coating in Fig. Supp. 6(e) was also directly extracted from the numerical solutions. The self-phase modulation (SPM) coefficients displayed in Fig. Supp. 6 (g) were then calculated according to the procedure outlined below. It can be seen that for longer wavelengths at $\lambda > 1450$ nm, the TMD's contribution to the SPM coefficients is larger than that of the silica core.

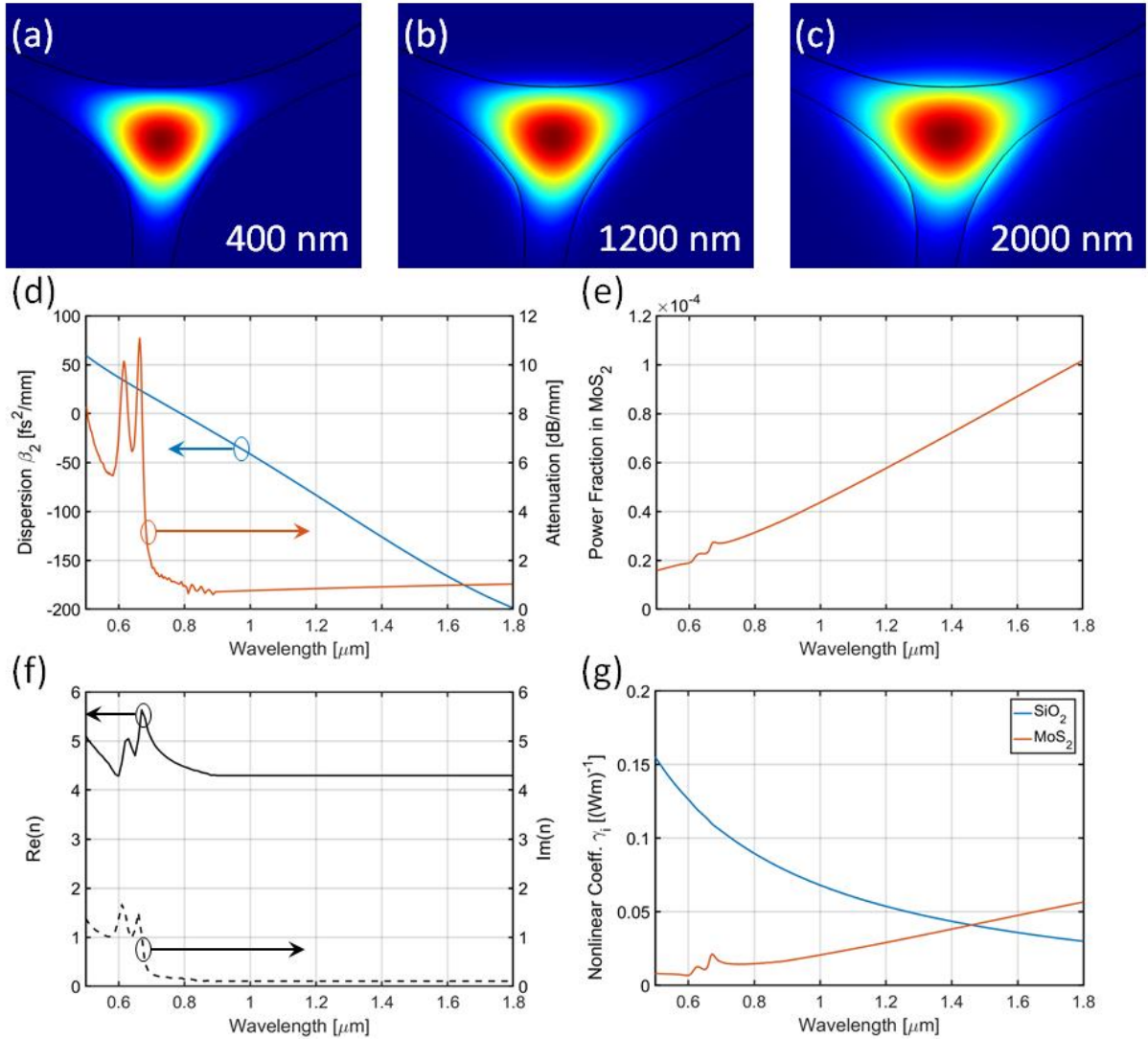

*Fig. Supp. 6: Properties of the fundamental mode on the ECF. (a-c) Absolute value of the electric field at wavelengths of (a) $\lambda = 400$ nm, (b) $\lambda = 1200$ nm, and (c) $\lambda = 2000$ nm. All modes are mostly polarized along the x-direction, i.e. parallel to the upper interface, which is coated with 0.65 nm of $MoS_2$. (d) Dispersion (blue) and attenuation (orange) of the $MoS_2$ coated ECF. (e) Fraction of the modal power flow (i.e. fraction of the integral over the z-component of the Poynting vector), occurring in the single layer of $MoS_2$. (f) Real (solid) and imaginary part (dashed) of the refractive index of $MoS_2$ (values are extrapolated and taken from [3]). (g) Contributions to the total nonlinear (i.e. self-phase modulation) coefficient $\gamma = \gamma_{SiO_2} + \gamma_{MoS_2}$ from each of the two materials, which have an overlap with the fundamental mode.*

*Tab. Supp. 1: Calculated Poynting vector z-component for two nondegenerate fundamental modes of the ECFs at $\lambda_0 = 1570$ nm.*

| Cross-Sectional part of the ECF | bare ECF x-pol. FM | bare ECF y-pol. FM | $MoS_2$ coated ECF x-pol. FM | $MoS_2$ coated ECF y-pol. FM |
|---|---|---|---|---|
| three air regions | 3.5 % | 3.3 % | 3.5 % | 3.3 % |
| air region above exposed surface only | 1.6 % | 0.9 % | 1.7 % | 0.9 % |
| monolayer | n/a | n/a | 0.008 % | 0.007 % |

## Supplement 7    Fiber Nonlinearity, Self-phase Modulation and Third Harmonic Generation Coefficients

TMDs exhibit very strong nonlinear optical effects [5] per unit thickness. On planar substrates, strong $\chi^{(2)}$-processes, i.e. second harmonic generation (SHG) and sum frequency generation have been investigated [6-9]. The former is further enhanced at edges [10], by exciton resonances [11, 12] and can be tuned by electronic doping [13] with external electric fields. The dependence of SHG on the number of monolayers and their orientation has been subject to intensive studies [14-16].

Third harmonic generation has also been investigated for TMDs. The efficiency of nonlinear light generation at the TH frequency is characterized by the nonlinear refractive index $n_2 = \frac{6}{8cn^2\varepsilon_0} Re(\chi^{(3)})$. For isolated single layer and multilayer $MoS_2$ crystals on planar substrates, values of $n_2^{MoS_2} \sim 30 \cdot 10^{-19}$ m$^2$/W have been found [17-21]. However, much higher values of $n_2^{MoS_2} \sim 2.7 \cdot 10^{-16}$ m$^2$/W have been reported on TMDs transferred on waveguides, indicating that the specific value of $n_2$ may depend on the geometry and substrate material [22], particularly in 2DFWGs. In our work we have used the latter, higher value, because of the similarity of the geometry in [22] with our ECFs and the consistency with our experimental findings.

Nonlinear effects in guided mode systems, however, do not only depend on the material nonlinearity but in the specific shape of modes and the distribution of electromagnetic energy over the nonlinear materials in the mode. Coefficients for Self-phase Modulation and THG have been calculated from overlap integrals, which take the highly vectorial nature of the mode into account.

For SPM we have resorted to the method developed in [23]. There are two distinct contributions to the overall SPM coefficient $\gamma$, which are related to the two sections of the geometry filled with SiO$_2$ and TMD, respectively. The coefficient is calculated from their sum, i.e. $\gamma = \gamma_{SiO_2} + \gamma_{MoS_2}$ and the contributions are computed according to:

$$\gamma_m(\lambda) = \frac{2\pi}{3\lambda} \frac{\varepsilon_0}{\mu_0} \Re(n^m(\lambda)) \, n^m(\lambda) \, n_2^m \frac{\iint_{A_m} \left(2|\mathbf{e}|^4 + |\mathbf{e}^2|^2\right) dA}{\left|\iint_{\mathbb{R}^2} (\mathbf{e} \times \mathbf{h}^*) \cdot \hat{z} \, dA\right|^2} \tag{1}$$

Where $m$ is the material index (SiO$_2$ or MoS$_2$), $A_m$ is the cross sectional area occupied by the material with index $m$ (i.e. either SiO$_2$ or MoS$_2$), $n^m(\lambda)$ and $n_2^m$ are material specific linear and nonlinear refractive indices, where the dispersion of the linear index is taken into account. $\mathbf{e}(x, y; \lambda)$ and $\mathbf{h}(x, y; \lambda)$ are the electric and magnetic field of a particular mode propagating inside the fiber at a given wavelength $\lambda$, and $\varepsilon_0$ and $\mu_0$ are the vacuum permittivity and permeability, respectively. $\hat{z}$ is the unit vector along the propagation direction.

We use the SPM coefficients to judge if the overall contribution of the TMD coating to the nonlinear effects shown in the fiber are substantial at all. Indeed, we find that the TMD coating has a significant impact on the ECF's nonlinear properties. At $\lambda_0 = 1570$ nm, we find that, the overall contribution of the TMD to the self-phase modulation coefficient $\gamma = \gamma_{MoS_2} + \gamma_{SiO_2}$ with $\gamma_{MoS_2}(\lambda_0) = 0.046$ (Wm)$^{-1}$ is slightly higher than that of $\gamma_{SiO_2}(\lambda_0) = 0.037$ (Wm)$^{-1}$ [23]. The evolution of both contributions as a function of the wavelength is given in Fig. Supp. 6 (g). Here we find that for longer wavelengths $\lambda > 1450$ nm, the TMD contribution is stronger than that of the silica core. For shorter wavelengths this is not the case. We attribute this to the contraction of the fundamental mode, as the wavelength is decreased and the consecutive reduction of the energy propagating in the TMD coating, as displayed in Fig. Supp. 6 (e). This means that the TMD coating may be used as a highly selective method to enhance the nonlinearities and tailor nonlinear 2DFWGs, while leaving dispersive properties of the 2DFWGs virtually unchanged.

The overlap coefficient of the THG is likewise a sum of the material specific contributions $\gamma_k^{(THG)} = \gamma_{SiO_2,k}^{(THG)} + \gamma_{MoS_2,k}^{(THG)}$, which is has an individual value depending on the choice of the higher order mode at the TH wavelength, here denoted by the arbitrary mode index $k$.

Here, the overlap coefficients are defined according to

$$\gamma_{m,k}^{(\text{THG})}(\lambda) = \sum_{ijnl} \frac{n_{2,ijnl}^{m} \iint_{A_m} e_{i,k}\left(\frac{\lambda}{3}\right) e_j^*(\lambda) e_n^*(\lambda) e_l^*(\lambda) dA}{\sqrt{\iint_{\mathbb{R}^2} e_{i,k} e_{i,k}^* dA \iint_{\mathbb{R}^2} e_j e_j^* dA \iint_{\mathbb{R}^2} e_n e_n^* dA \iint_{\mathbb{R}^2} e_l e_l^* dA}} \qquad (2)$$

Where the subscripts $ijnl$ denote the polarization direction of the electric field and $e_{i,k}$ is the electric field of the third harmonic higher order mode with mode index $k$ and $e_j$, $e_n$, and $e_l$ denote the fundemntal mode at the fundamental wavelength. Because the fields are mostly polarized in $x$ and $y$-direction and there is no appreciable value for the out-of-plane nonlinear tensor of the TMD, the equation simplifies to:

$$\gamma_{m,k}^{(\text{THG})}(\lambda) = \frac{n_{2,xxxx}^{m} \iint_{A_m} e_{x,k}\left(\frac{\lambda}{3}\right) e_x^*(\lambda) e_x^*(\lambda) e_x^*(\lambda) dA}{\sqrt{\iint_{\mathbb{R}^2} e_{x,k}\left(\frac{\lambda}{3}\right) e_{x,k}^*\left(\frac{\lambda}{3}\right) dA} \left(\iint_{\mathbb{R}^2} e_x(\lambda) e_x^*(\lambda) dA\right)^3} \qquad (3)$$

## Supplement 8  Laser Pulse Characteristics for the Excitation of Third Harmonic

The pulse duration was estimated by measuring the spectrum and the non-collinear intensity autocorrelation (AC) of the pulse incident to the fiber (see Fig. Supp. 7 (a) and (b)). The measured AC was compared to the AC of the transform-limited pulse displayed in Fig. Supp. 7 (c) as calculated from the measured spectrum. Both are found in excellent agreement in the central peak region as well as in good agreement in the wings. We conclude that the FWHM pulse duration is close to the transform-limit of 30 fs. The final estimate for the pulse duration of 32 fs at the fiber facet was obtained by further numerical propagation through the used aspheric lens.

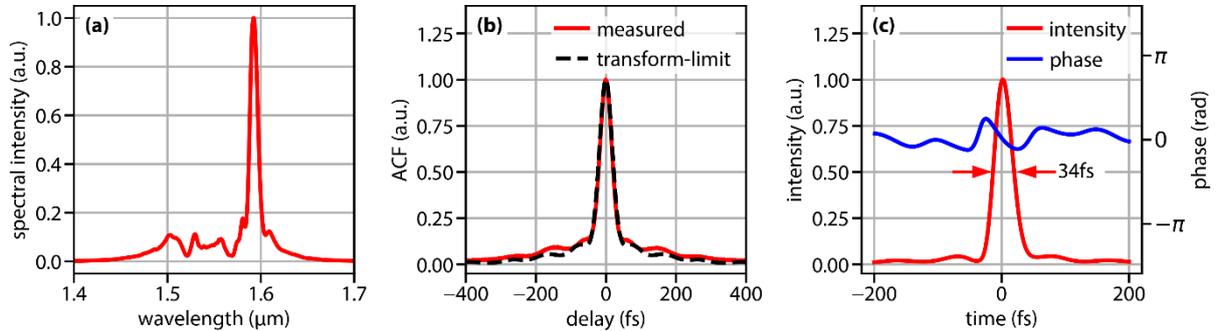

*Fig. Supp. 7: Laser pulse characteristics. (a) Laser spectrum at the input of the fiber. (b) Measured intensity autocorrelation (red) compared with the autocorrelation of the transform-limited pulse. (c) Transform-limited pulse as calculated from the spectrum propagated numerically through the fiber-coupling optics.*

## Supplement 9  HOM Third Harmonic Generation

Third harmonic generation (THG) was excited by a femtosecond laser emitting at $\lambda_0 = 1570$ nm (center of mass of the spectrum). It was observed with a spectrometer according to the procedure described above. Spectra for three different input power levels, for a bare and an MoS₂ coated ECF are displayed in Fig. Supp. 8 (a) and in the main text. For the MoS₂ coated ECF we have inferred the input power from the output power and the previously inferred loss per unit length of the fiber. This means that in fact for the MoS₂ coated ECF there is less pulse energy available to drive the THG process as it continuously drops along the length of the fiber. Nevertheless, we observe a 47 % increase in THG for an input energy of 0.59 nJ, 41 % for 0.67 nJ and 18 % for 0.70 nJ input energy, as indicated by the ratio $\epsilon_{\text{THG}} = E_{\text{MoS}_2}^{\text{THG}} / E_{\text{Bare}}^{\text{THG}}$, displayed in the first row of Fig. Supp. 7 (c).

The TH light does not correspond with a third of the wavelength of the FW. In fact, most of the output spectrum is centered at wavelengths between 540 and 560 nm, which corresponds to a FW wavelength of 1620 to 1680 nm. Both ranges are marked with orange bands in Fig. Supp. 8 (a) and (b). In this spectral range some higher order modes of the TH wavelength are phase matched with the FM, i.e. if the effective refractive index of the FW mode $n^{\text{FW}}$ and of the TH mode $n_k^{\text{TH}}$, fulfils the inequality $\Delta n = |n^{\text{FW}} - n_k^{\text{TH}}| < \lambda_0/(\pi L_c)$. Here $k$ is the number of the higher order mode (HOM). $L_c$ is the characteristic length of the system, i.e. the typical length of an MoS$_2$ crystal, which is $29\ \mu m$ (Fig. Supp. 5), yielding $\Delta n = 0.012$. For our ECFs only HOMs with $k \gg 1$ are phase matched. Note that we have ignored the added contribution of the broadband nature of the excitation pulse to the PM-bandwidth as it is small, if compared to the value coming from the phase-matching length.

Because of this mismatch of laser wavelength and PM band, any THG must be precipitated by a spectral broadening of the FW into this band, via nonlinear processes, which explains the higher than cubic power scaling of the THG process displayed in the inset of MoS$_2$ coated Fig. Supp. 8 (a).

The broadening process of the FW into the THG relevant sub-band differs somewhat for the MoS$_2$ coated and bare ECFs. While a more systematic investigation of spectral broadening in MoS$_2$ coated ECFs is subject to further investigation, we here resort to estimate its impact on the THG efficiency only. We therefore measure the energy in the THG generation sub band of the FW and compare these values between MoS$_2$ coated and bare ECFs. The corresponding ratio $\epsilon_{\text{FW}}$ is calculated as $\epsilon_{\text{FW}} = \int_{1620\ \text{nm}}^{1680\ \text{nm}} E_{\text{Bare}}(\lambda)d\lambda / \int_{1620\ \text{nm}}^{1680\ \text{nm}} E_{\text{MoS}_2}(\lambda)d\lambda$. This ratio is displayed in the center row of Fig. Supp. 7 (c), and we get $\epsilon_{\text{FW}} = 0.96$ for $E = 0.59$ nJ, $\epsilon_{\text{FW}} = 1.03$ for $E = 0.67$ nJ and $\epsilon_{\text{FW}} = 1.12$ for $E = 0.70$ nJ.

Using these two ratios and the fact that the THG efficiency scales with the cube of the FW energy and linearly with the TH energy, we can estimate the relative efficiency of the THG process of the MoS$_2$ coated ECFs vs. the bare ECFs as $\epsilon_{\text{THG}} \cdot \epsilon_{\text{FW}}^3$. The results are displayed in the bottom row of Fig. Supp. 7(c). They indicate a corrected enhancement factor of 30% for $E = 0.59$ nJ, 53% for $E = 0.67$ nJ, and 67% for $E = 0.70$ nJ, respectively. This corrected enhancement factor is consistently higher than the uncorrected one and underlines that the THG process is fundamentally enhanced by the TMD coating.

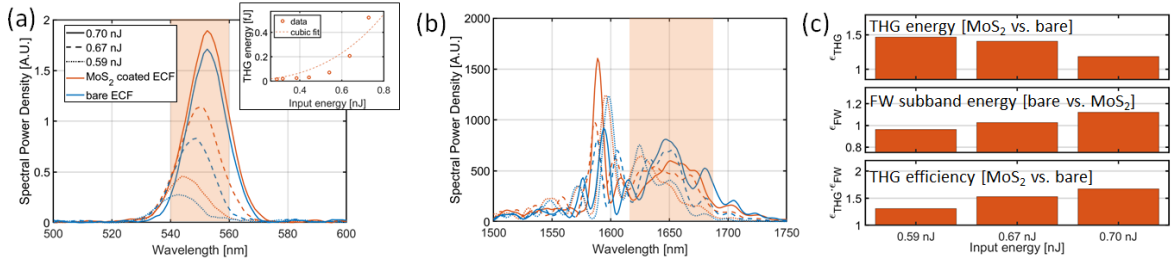

*Fig. Supp. 8: Third harmonic and infrared spectra for MoS$_2$ coated and bare ECFs at different input energies. (a) Third harmonic spectra. The THG band from 540 to 560 nm is marked in orange. (Inset) Power dependence of the THG for the MoS$_2$ coated ECF, with a cubic fit. (b) Nonlinearly broadened fundamental wave spectra. The orange shading specifies the THG relevant sub-band indicated by the equivalent shading in (a). Same legend as (a). (c) Comparison of THG energy (upper), FW energy in the THG relevant sub-band (center) and estimation of THG efficiency (lower) of MoS$_2$ coated ECFs vs. the bare ECFs as a function of the input energy.*

The value of the overlap coefficient for the THG process depends critically on the shape of the specific HOM mode, or more specifically their x-polarized components. In the phase matched regions we have identified three modes which have a large overlap coefficient and which contribution to the experimentally recorded picture of the TH modes.

The modes have been designated as $M_1$, $M_2$ and $M_3$ in the main text. Their shape, determined from the FEM simulations discussed above, is displayed below in Fig. Supp. 9 (a-c). Each of the modes does have at least one field maximum close to the center of the top surface of the ECF. It is thus obvious that each of these three mode's overlap coefficients does grow considerably upon the application of the TMD coating.

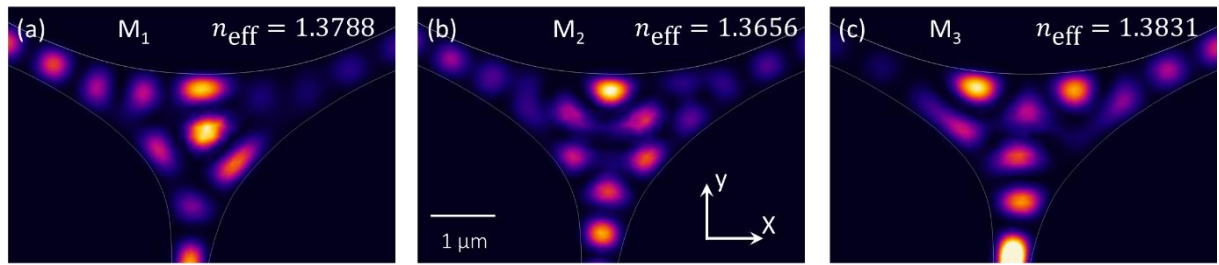

*Fig. Supp. 9: $|E_x|^2$ of the three fundamental modes $M_1$, $M_2$, and $M_3$ which have been determined to be crucial for the THG-process of the ECFs. (a) $M_1$ at $n_{\text{eff}} = 1.3788$, which is the dominant mode for the bare and $MoS_2$-coated ECF. (b) $M_2$ at $n_{\text{eff}} = 1.3656$, which has the second largest overlap coefficient for the bare ECF. (c) $M_3$ at $n_{\text{eff}} = 1.3831$, which has the third largest overlap coefficient for the $MoS_2$ coated ECF.*